\documentclass[aps,prl,twocolumn,epsf,epsfig,amsmath]{revtex4}

\usepackage{latexsym}
\usepackage{hyperref}
\usepackage{times}
\usepackage{graphicx}
\usepackage{amsmath}
\usepackage{multirow} 
\usepackage{amsthm}
\usepackage{dcolumn}
\usepackage{bm}
\usepackage{ textcomp }
\usepackage{color}
\usepackage{amssymb}
\usepackage{xcolor}
\usepackage{easyReview}
\usepackage{mathdots}
\usepackage{float}
\usepackage{lineno}
\usepackage{todonotes}
\usepackage{array}

\newcommand{\beq}{\begin{eqnarray}}
\newcommand{\eeq}{\end{eqnarray}}

\begin{document}
\title{The Luttinger Count is the Homotopy not the Physical Charge: Generalized Anomalies Characterize Non-Fermi Liquids}
\author{Gabriele La Nave$^\dagger$, Jinchao Zhao$^*$, and Philip W. Phillips$^*$}

\affiliation{$^\dagger$Department of Mathematics, University of Illinois at Urbana-Champaign, Urbana, IL 61801, USA}
\affiliation{$^*$Department of Physics and Institute of Condensed Matter Theory, University of Illinois at Urbana-Champaign, Urbana, IL 61801, USA}

\begin{abstract}
We show that the Luttinger-Ward functional can be formulated as an operator insertion in the path integral and hence can be thought of as a generalized symmetry.  The key result is that the associated charge, always quantized, defines the homotopy, not the physical charge.   The disconnect between the two arises from divergences in the functional or equivalently zeros of the single-particle Green function.  Such divergences produce an anomaly of the triangle-diagram type.  As a result of this anomaly, we are able to account for the various deviations\cite{rosch,dave,altshuler,osborne,l3,l5} of the Luttinger count from the particle density.
As a consequence, non-Fermi liquids can be classified generally by the well known anomaly structures in particle physics.  Charges descending from generalized symmetries, as 
in the divergence of the Luttinger-Ward functional, are inherently non-local, their key experimental signature. 

\end{abstract}
\date{December 2024}

\maketitle

Schwinger terms\cite{schwinger,schwinger2} point to an anomaly as they arise from regularizing a quantum field theory.  In the path integral, this typically originates from a symmetry that does not preserve the measure, thereby leading to an anomalous Ward identity, such as the axial symmetry in QED where the divergence of the axial $U(1)$-current is non-zero (and equals the norm square of the Gauge field strength)\cite{schwinger,schwinger2}.  Indeed, anomalies for continuous symmetries have helped the resolution of fundamental questions in QCD where chiral symmetry is preserved in the isospin sector but broken spontaneously in the isospin singlet channel leading to decay of $\pi^0$ to two photons, the ABJ anomaly\cite{adler, bell-jackiw}. While the scattering matrix for this decay process can be shown to be non-zero from the triangle diagrams\cite{adler, bell-jackiw}, the crucial aspect\cite{ginsparg,nielsen} that leads to the ABJ anomaly is captured entirely by the Atiyah-Singer (AS)\cite{asindextheorem} index theorem. Consequently, the ABJ anomaly is rooted in topology and hence invariant under renormalization.  

However, not all anomalies have a clear connection to the AS index theorem. 't Hooft\cite{thooft} anomalies arise upon gauging a global symmetry, rather than a local one as in ABJ, of the underlying quantum field theory.   Augmenting the gauged quantum theory with auxiliary fermionic fields whose sole purpose is to eliminate the apparent ABJ anomaly lays plain\cite{thooft} an equivalence between the UV and IR anomalies.

Indeed, it is the primacy of anomalies in particle physics, that has led to their sightings in the consideration of non-Fermi liquids\cite{altshuler,bc,else2}.  Ref.\cite{else2} derives the invariance of the Fermi surface to interactions, the Luttinger  count\cite{luttinger}, (Eq. (24) of their paper\cite{else2} ) from Eq. (19) (of their paper) which directly follows from the t'Hooft anomaly.  Heath and Bedell\cite{bc} find a relationship between the AS index theorem and the validity of the Luttinger count. Alternatively, Altshuler and colleagues\cite{altshuler} find a disconnect between the Luttinger count and the charge density stemming from the same mechanism in the triangle diagrams in the ABJ chiral anomaly.  Namely, an integral\cite{altshuler} that is supposed to vanish by symmetry fails to do so because it contains a linear divergence.  Regularization of this divergence leads to a splitting of a double pole into two poles that lie in opposite half-planes.  It is the splitting of the poles that leads to the deviation of the Luttinger count from the particle density. We note that it is precisely the splitting of the double pole to two poles that is the mechanism for Mott insulation\cite{dzy,phillipsrmp} and the onset of zeros in the single-particle Green function. 

These different approaches to positioning the Luttinger count in the context of anomalies suggest that a more in-depth analysis of this possibility is warranted. While more restrictive statements of the Luttinger count exist (which are applicable only to gapless systems\cite{oshikawa}), its most general form\cite{luttinger},
\beq
n=2\int_{\rm Re G(k,\omega=0)>0}\frac{d^Dk}{(2\pi)^{2D}},
\label{lutt}
\eeq
equates sign changes of the real part of the single-particle Green function, $G(k,\omega)$, with the particle density, $n$. Clearly, the RHS of Eq. (\ref{lutt}) must be an integer. However, numerous studies abound\cite{rosch,dave,osborne,l3,l5}  which demonstrate that the RHS and LHS of Eq. (\ref{lutt}) are not equal. Hence, the equation is not valid. If sign changes of $G$ do not enumerate the physical charge density, then what do they count?  Should an anomaly lurk in the derivation of Eq. (\ref{lutt}), then it is rooted in the structure of the underlying Luttinger-Ward (LW) functional\cite{lw}. Thus far, no such unpacking of Eq. (\ref{lutt}) and anomalies  has emerged. 
For example, while the ABJ anomaly requires the identification of a symmetry that quantization destroys, none was noted in \cite{altshuler}.  The 't Hooft anomaly work\cite{else2,else1} is based entirely on the bulk-boundary correspondence where the bulk is represented by a $D+1$-dimensional Chern-Simons action.  This is of course true only if the symmetry group is a connected Lie group.  For a disconnected or a finite symmetry group, the Wess-Zumino constraints\cite{wz} no longer apply\cite{kapustin}, thereby invalidating the correspondence with 't Hooft.  We bring this up because Haldane and Anderson\cite{ah} have shown that the ultimate symmetry group for a $3D$ Fermi liquid is the disconnected group $O(4)$ and hence lies outside the analysis in \cite{else2}.  Additionally, underlying the correspondence\cite{else2} is the assumption that an anomaly inflow from the bulk cancels the 't Hooft anomaly in the boundary.  Ultimately, this inflow is governed by the Wess-Zumino consistency conditions which obviously fail for generalized symmetries for which only a Lie algebroid, at best, exists.

What we show here is that a generalized symmetry inherent in the Luttinger-Ward functional\cite{lw} is such that when broken, in the t'Hooft sense, a non-Fermi liquid arises in which the Luttinger count fails.   By generalized symmetry, we mean then a symmetry of an extended path integral defined not just on fields over the underlying space-time but over the extended space that includes the loop space $\Omega_1$. 
Concretely, the mechanism is rooted in the appearance of zeros of the single-particle Green function which have been shown previously\cite{rosch,dave} to a breakdown of Eq. (\ref{lutt}).  From our work, we are able to make a connection with the von Neumann algebra interpretation of generalized symmetries due to Casini\cite{casiniarxiv}. 



The celebrated Luttinger count\cite{luttinger} relies on a perturbative proof which soundly depends on Fermi's golden rule and the analyticity of the self energy near the Fermi surface. In general terms, under the assumption of analyticity of the self-energy,  one can reduce Luttinger's count to the validity of the following identities 
\begin{equation}
    \begin{aligned}
      \quad &I_1\equiv \frac{i}{2\pi} \int \; \frac{d^nk}{(2\pi)^n}\, \int_\gamma d\omega \; \frac{\partial }{\partial \omega} \log G(\omega, k)= \frac{N}{2V},\\
      \quad &I_2\equiv -i\int \; \frac{d^nk}{(2\pi)^n}\, \int_\gamma \;\frac{d\omega}{2\pi} \left(G(\omega,k) \frac{\partial \Sigma(\omega,k)}{\partial \omega} \right)=0,\\
       \end{aligned}
    \label{lutt-integrals}
\end{equation}
where $\gamma$ is a path counter-clockwise enclosing the real axis.
We now recall the reason why the integral $I_2$ vanishes, under the assumption of analyticity of $\Sigma$. First, it is well understood by now that the validity of the Luttinger count for an interacting fermionic system is intimately related to the analyticity\cite{dave} of the LW functional $\Phi[G]$. Defined\cite{lw,lw1} as the sum of all skeleton bubble diagrams, the LW functional ($\Phi[G]$) is related to the  Kadanoff-Baym\cite{kb} functional, the two particle irreducible effective action, through 
\begin{equation}
   \Gamma[G]=\Phi[G]-\frac12\left(Tr[\Sigma \, G]+Tr\log(-G)\right).
\end{equation}
As shown by Luttinger and Ward\cite{lw}, $\Phi[G]$ satisfies the differential equation,
\begin{equation}\label{LW-eq}
   \frac{\partial}{\partial G_{\alpha \beta}(\Vec{k}, i\omega _\nu)} \Phi[G]= \frac{1}{\beta} \Sigma _{\alpha \beta}(\Vec{k}, i\omega _\nu).
   \end{equation}
This equation is equivalent to the statement that $\Sigma \delta G$ is an exact differential, provided it is analytic (at least near the fermi surface) 
\begin{equation}
    \delta \Phi[G]= \frac{1}{V} \; \sum _{k,\sigma} \int \frac{d\omega}{2\pi} \Sigma _\sigma (\omega, k)\delta G_\sigma,
    \label{exact}
\end{equation}
where $\sigma$ are the spin indexes. Equation \eqref{exact} implies the vanishing of integral $I_2$ in \eqref{lutt-integrals}, after a mere integration by parts. In\cite{dave}, they show that the $SU(N)$-atomic Hubbard model fails the Luttinger count precisely because of the non-analyticity of the self-energy.

In order to understand why the Luttinger count boils down to the validity of the equations for $I_1$ and $I_2$, we reformulate the proof slightly differently. It is well known that if $J$ and $K$ are, respectively, the ~1-particle and ~2-particle sources, then the partition function $Z[G]=exp \{ iW(J,K)\}$ is equal to 
\begin{equation}\label{partition}
    Z[G]= \bar Z \, \exp \{ \Phi[G]-\frac12\left(\rm Tr[\Sigma \, G]+Tr\log(-G)\right)\} .
\end{equation}
Using the Dyson equation, one can see that the self-energy, 
$\Sigma=G_0^{-1}-G^{-1}$ and therefore $\Sigma G=(G_0^{-1}-G^{-1})G$,
and in particular 
\begin{equation}\label{anom1}
    W\propto \Phi[G]-\frac12\left(\rm Tr[(G_0^{-1}-G^{-1})G]+Tr\log(-G)\right).
\end{equation}
One derives from \eqref{LW-eq} that 
$\Sigma \delta G=(G_0^{-1}-G^{-1})\delta G=  G_0^{-1}\delta G- \delta \log G$ and therefore that
\begin{equation}\label{fi}
    \Phi=\frac{1}{\beta} \int d\omega \left( G_0^{-1}\delta G- \delta \log G\right), 
\end{equation}
where $G_0$ is the bare (non-interacting) Green function, and therefore $G_0^{-1}$ is analytic throughout the complex plane. 
Whence one clearly sees that\cite{dave}
\begin{equation}
    n=I_1+I_2,
\end{equation}
where $I_2$ represents the failure of the theorem.
The term $I_2$ 
vanishes when $\Phi[G]$ is analytic since by Eq. \eqref{exact} $\Sigma \delta G$ is exact ($\delta \Phi= \int \Sigma \delta G$).
But from equation \eqref{fi}, we see that $\Phi[G]$, as a result of the $\ln G$ dependence, cannot be analytic in the vicinity of a Luttinger surface, although by the analyticity of $G$ near the Luttinger surface, the singularities of $\Phi[G]$ are not essential (but are rather those of a meromorphic function). By similar reasoning, we recast the first part of the Luttinger count as $I_1$ in Eq. (\ref{lutt-integrals}). 


Rephrasing the key observation of generalized symmetries\cite{ortiz,gksw}, we note that a classical symmetry represented by a group preserving the action, when non-anomalous (i.e. when the action of the path-integral is preserved as well), should be thought of as an operator that can be inserted into the path integral, along sub-manifolds (of spacetime) of non-negative codimension. These operators should be invariant under continuous deformations of the submanifolds (i.e., they should only depend on the homology class of the submanifold).   
For a standard example of how to view the symmetry of a QFT in this fashion, one can consider the current $j$ in QED, where the classical conservation equations are $d\star j=0$ (where $\star$ is the Hodge star operator). To any closed codimension ~1 submanifold, $\Gamma$, we can associate the global charge $Q(\Gamma)= \int _\Gamma \, \star j$ which can be used to construct an operator that resmbles the well-known Wilson ``line" (here $\Gamma$ is a codimension ~1 object, a Wilson line integrates over a curve)
\begin{equation}
    W_\Gamma(j, g=e^{i\alpha})= \mathrm{exp} ( i\alpha Q).
\end{equation}
By Stokes' theorem, $Q(\Gamma)$ is a topological operator, i.e., it only depends on the homotopy class of $\Gamma$.
Taking N\"other's lead (her first theorem) in which a continuous symmetry is associated with a conserved current, we regard the Wilson loop as a generalization of this idea as has been proposed previously\cite{mcgreevy,ortiz}.  Namely, we interpret the operator obtained by integrating a closed p-form over a p-dimensional submanifold as a generalized symmetry. The charges that result from this operation have no foundation in a classical symmetry.  We therefore view a generalized symmetry\cite{ortiz,casini21} as a Wilson loop-like operator of which the Luttinger-Ward functional is an example. One of our contributions is to identify the action of a groupoid as being associated with generalized symmetry or homotopy charges.

Another important and elucidating example of such generalized symmetries arises in the string context. In \cite{vafa}, they are concerned with a 2D theory which is gauged by a finite group $G$. If $\hat G=\{ \text{irr. rep.'s } \rho:G\to U(1)\}$ is its Pontryagin dual, then $G$ and $\hat G$ are non-canonically isomorphic.  Gauging by a finite group is equivalent to considering a $G$-orbifold over the original 2D manifold.  
A $G$-orbifold has a dual $\hat G $ -symmetry assigning charges to twisted sectors, and orbifolding again by this dual $\hat G$ symmetry recovers
the original theory(\cite{vafa}). The dual $\hat G$ symmetry is known as the quantum symmetry\cite{vafa}. The fact that in the dual twisted sector, one obtains charges demonstrates the nature of the generalized symmetries as being akin to symmetries. 
If the group $G$ is non-abelian, the duality can still be expressed, but one needs to make use of generalized symmetries \cite{finite}.

Thus, the crucial observation is that generalized symmetries do not have to arise from group actions. 
 Rather, they only depend on operator insertions. 
For our purposes, we consider the Luttinger-Ward operator $e^{\Phi[G]}$, which because of Eq. \eqref{partition}, we think of as an operator insertion in the path integral obtained by $\bar Z \, \exp \{ -Tr[\Sigma \, G]+Tr\log(-G)\}$.  Because generalized symmetries are associated only with charges defined by the homotopy class, $Q$ should be properly thought of as a homotopy charge, not the physical charge.  As we will see, this distinction is crucial because the proper quantity defined by the LW functional is the homotopy, not the physical charge.  It is this twist that has led to the myriad of results for the Luttinger count\cite{rosch,dzy,volovik,l3,osborne,l5}.  Now, in the presence of zeros, our claim is that generically the Wilson-like operator $e^{\Phi[G]}$, which by Eq. \eqref{fi} we can think of as a generalized symmetry, identified by $\Phi[G]$ (see Eq. \ref{exact}), gives rise to an anomaly of the ABJ kind.


Generalized symmetries arise naturally in two ways as generalizations of symmetries of the Hamiltonian. One way is the fact that smooth homotopies (i.e. isotopies) of a path as in a 1-dimensional sub-manifold arise naturally from the local action of a vector field which generates the "motion" of the path. 
 These actions arise then from a local $S^1$ action on the space of paths. The other way is intimately related (and in some sense equivalent) to the recent realization in QFT and quantum gravity \cite{casini22} that "conserved" currents are a natural manifestation of the presence of non-local operators in the local algebras of regions of spacetime. Our observation differs from the aforementioned approach in that we focus on the algebras of operators localized in regions in momentum space (rather than spacetime) and more specifically we use the Fermi surface and the Luttinger surface as natural boundaries, and in particular we focus on regions separated by such surfaces (rather than by causality).
The significance of the Luttinger surface in particular lies in the fact that it is the culprit for the presence of the singularities of $\Phi[G]$. 

First, let us explain in what sense a homotopy of a path $\gamma$ (or of any $p$-cycle for that matter) can be thought of as a form of symmetry transformation of the spacetime. A homotopy between two paths $\gamma_1$ and $\gamma _2$ in some space $M$is a continuous map $\gamma : I\times I \to M$ such that $\gamma (0,x)= \gamma_1(x)$ and $\gamma(1,x) = \gamma _2(x)$. Such continuous maps can always be approximated by smooth maps and so we assume the map $\gamma$ to be smooth.  On the surface swept out by $\gamma (t,x)$, the {\it motion} from $\gamma_1(x)$ to $\gamma _2(x)$ can be obtained by means of the integral curves of a vector field $V$ defined on the surface. We can extend from zero $V$ to a neighborhood of $\gamma$ in the whole space, and this vector field (which we still call $V$ by abuse of notation) generates a transformation (a diffeomorphism) which maps (at least locally) $\gamma _1$ to $\gamma _2$. 
We argue that these local transformations {\begin{footnote}{Technically, what we are defining gives rise to the action of {\it groupoid}, not a group}\end{footnote}} are the ones that experience the anomaly. To make this discussion more precise, we define the extended space-time as the space $M\times \Omega_1(M)$, where $\Omega_1(M)=\{ \gamma :[0,1]\to M,\;\; \gamma(0)=\gamma (1)\}$, so $\pi_1(M)$, the first fundamental group is nothing but the group of loops modulo homotopy equivalence or $\pi_1(M)=\Omega ^1(M)/\equiv$, where $\equiv$ is understood as homotopy equivalence.  Given any (non-necessarily smooth)1-form $j$ (such as a current), we then define the new path integral
\begin{equation}\label{loop-ext}
    \mathcal Z= \int_{F\times \Omega ^1(M)} \mathcal D \phi \mathcal D \gamma e^{iS(\phi)+i \int_\gamma j },
\end{equation}
where $F$ is the bundle whose sections define the fields of the given theory (determined by a path integral $Z= \int_{F} \mathcal D \phi \mathcal D \gamma e^{iS(\phi)}$).
Clearly, if $j$ is conserved, $\int_\gamma j$ is only dependent on the homotopy class $[\gamma]$ of the path and therefore $\mathcal Z$ descends to a path integral on $M\times \pi _1(M)$,
\begin{equation}\label{pi1-ext}
     Z_{ext}= \int_{F\times \pi_1(M)} \mathcal D \phi \mathcal D \gamma e^{iS(\phi)+ i\int_\gamma j } ,
\end{equation}
which we can interpret as a charge-dressing of the original theory $Z$ (here we use a regularization of the measure $\mathcal D \gamma$). Equivalently, Eq. (\ref{pi1-ext}) follows from Eq. (\ref{loop-ext}) only if the partition function, ${\mathcal Z}$ is anomaly free in which case $\int_\gamma j$ is the original charge.  While we have constructed this argument entirely for a 1-dimensional path, it is easily extendable to higher-dimensional currents, $j$, simply replacing the first homotopy group with a higher homology group.

The Luttinger surface $S_L$ has a trivial normal bundle and hence has the form $S_L\times \mathbb R^k$, where $k$ is the codimension of $S_L$.  Then a neighborhood of thickness $\Lambda$ of $S_L$ is diffeomorphic to $S_L\times [-\Lambda , \Lambda]$. If we consider a path entirely contained in such a neighborhood, then we can think of a smooth homotopy between two loops as arising from the local action of a group-like object (a groupoid) and this action, due to the presence of the singularity of $\Phi[G]$, represents an anomaly of the measure. 
Observe that $e^{i\Phi[G]}$ can be thought of as an invertible field theory.

An alternate explanation of the generalized symmetry associated with $\Phi[G]$ stems from \cite{casini22}. In \cite{casini22} and references therein, the notion of generalized symmetry\cite{ortiz} comes about when considering the net of local algebras of observables (operators) of the theory. Usually the net of algebras (which have to be, for physical reasons, von Neuman algebras) is constructed by associating an algebra of operators $\mathcal A(R)$ with every spacetime region $R$ which satisfies some properties, notably, {\it additivity} $\mathcal A(R_1\cup R_2)= \mathcal A(R_1) \bigvee \mathcal A(R_2)$ and {\it isotonia} $\mathcal A(R_1)\subseteq \mathcal A(R_2)$ if $R_1\subseteq R_2$) and {\it causality} $\mathcal A(R) \subseteq \left(\mathcal A(R')\right)'$ (here $R'$ means {\it causal complement of } $R$ and $\mathcal A'$ denotes the commutant of $\mathcal A$, that is the algebra of all bounded operators which commute with $\mathcal A'$.
The novel understanding from \cite{casini22,casini21,casiniarxiv} is that for some region $R$, the von Neumann algebra 
$\mathcal A(R)$ associated with $R$, instead of being equal to the von Neumann dual of $\mathcal A(R')$, might actually be strictly contained in $\mathcal A(R')'$ (it's always contained). That is,
\beq
  \mathcal A(R')'= \mathcal A(R)\bigvee \{ a\}  
\eeq
where $\{ a\} $ denotes a set of operators (observables) that commute with the local observables in $R$ (i.e., $\mathcal A(R)$) but are not contained in $\mathcal A(R)$. This is because by definition $\mathcal A(R')'$ are all the operators that commute with $\mathcal A(R')$.
Therefore, there exist "non-locally generated operators" associated with the region $R$. These are the generators of the generalized symmetries, such as the Wilson line denoted above.  
Observe that in the presence of zeroes, the Luttinger-Ward functional is inherently not perturbative, as can be seen for instance by the fact that the Green function has two poles which split.  Therefore, $\Phi[G]$ has to contain a non-locally generated operator in the sense above and therefore it represents a generalized symmetry as defined by \cite{casini22,casini21,casiniarxiv}.  

A question arises: Why does the perturbation theory break down.  For the case of the bifurcation of the poles, it arises because part of the $O(4)$\cite{ah} group is broken by the interactions.  As pointed out previously\cite{discrete} it is the $Z_2$ symmetry of a Fermi surface that is broken by the interaction.  As this is a discrete symmetry, the Wess-Zumino conditions for anomalies are vacuous\cite{kapustin}.  That is, the previous analysis focused on 't Hooft anomalies\cite{else1,else2} and the Luttinger count is inherently insensitive to the existence of this symmetry.


Without losing generality, we consider a single-particle Green function which supports a bipartite spectral weight as suggested by the splitting of the double pole to two poles\cite{altshuler}. Such a Green function follows necessarily\cite{discrete} from breaking the discrete $Z_2$ symmetry\cite{ah} of a Fermi liquid. The Green function is given by,
\begin{equation}
    G(z,k)=\frac{Z_1}{z-\xi(k)-U_1(k)}+\frac{Z_2}{z-\xi(k)-U_2(k)},
    \label{eq:green2p}
\end{equation}
where $Z_1$ and $Z_2$, satisfying $Z_1+Z_2=1$, are the spectral weights of the 2 poles of the Green function,  $\xi$ is the dispersion of the non-interacting system, and $\xi+U_1<\xi+U_2$ are the locations of the two poles.
The self energy $\Sigma(z,k)$ is defined
\begin{equation}
\begin{split}
    \Sigma(z,k)&=G_0^{-1}(z,k)-G^{-1}(z,k)\\
    &=Z_1U_1+Z_2U_2+\frac{Z_1Z_2(U_1-U_2)^2}{z-\xi-Z_1U_2-Z_2U_1}.
\end{split}
\end{equation}
The self-energy supports a simple pole at $z=\xi+Z_1U_2+Z_2U_1$. Such a divergence leads to a zero of the single-particle Green function.

All we have to do to obtain the charge density, $n=I_1+I_2$, is evaluate the integrals for $I_1$ and $I_2$ detailed in the Appendix.  Let us define $a_{12}=n_F(\xi+Z_1U_2+Z_2U_1)$. We compute the integrals (see Appendix) in $I_2$ using the residue theorem 
which leads to the final result
\beq
I_2=-Z_2n_F(\xi+U_1)+a_{12}-Z_1n_F(\xi+U_2).
\eeq
Several cases arise for the location of the poles and zeros.
\begin{enumerate}
    \item Both poles are located on the left-half-plane, which means $n_F(\xi+U_1)=n_F(\xi+U_2)=n_F(\xi+Z_1 U_2+Z_2 U_1)=1$ . There are in total 3 poles or zeros located on the left half plane leading to $I_1=1$. Thus the difference from Luttinger count yields
\beq
    I_2=-Z_2+1-Z_1=0,
\eeq
implying the Luttinger count equals the particle density.
\item The two poles located one on either side of the imaginary axis, and the zero is on the left-half-plane.  This means that $n_F(\xi+U_1)=n_F(\xi+Z_1U_2+Z_2U_1)=1$ and $n_F(\xi+U_2)=0$. There are in total 2 poles or zeros located on the left-half-plane, implying $I_1=0$. Thus the difference from the charge density
\begin{equation}
    I_2=-Z_2+1=Z_1
\end{equation}
is non-zero and the Luttinger count gives the \emph{wrong} answer.
\item The two poles located on both sides of the imaginary axis, and the zero is located on the right-half-plane, signifying that $n_F(\xi+u1)=1,n_F(\xi+u2)=n_F(\xi+Z_1U_2+Z_2U_1)=0$. Just a single pole or zero is located on the left-half-plane, leading to $I_1=1$. Thus the difference from Luttinger count is
\begin{equation}
    I_2=-Z_2.
\end{equation}
non-zero and consequently, the Luttinger count supplies the \emph{wrong} answer.
\item Both poles are located on the right-half-plane, which means $n_F(\xi+u1)=n_F(\xi+u2)=n_F(\xi+Z_1U_2+Z_2U_1)=0$. There are in total 0 poles or zeros locate on the left-half-plane, $I_1=0$. Thus the difference from Luttinger is
\begin{equation}
    I_2=0
\end{equation}
implying the Luttinger count is identically the particle density.

\end{enumerate}

Based on the above analysis, the Luttinger count does not coincide with the occupancy once there are both positive and negative components to the spectral weight, the source of zeros of the single-particle Green function.  This naturally obtains in a Mott insulator\cite{phillipsrmp} and any bi-partite Green function\cite{l3}. Even in such cases, the homotopy charge is still quantized but just disconnected from the physical charge.  This disconnect arises from an anomaly in the path integral. The anomaly is precisely of the ABJ\cite{adler,bell-jackiw} type as suggested previously\cite{altshuler}. 't Hooft anomalies are included in this picture as they become triangle-diagram anomalies only upon gauging, that is going from Eq. (\ref{loop-ext}) to Eq. (\ref{pi1-ext}).  Quite generally,  non-Fermi liquids must arise from an anomaly.  Without it, the LW functional is completely well behaved and the homotopy and physical charges coincide.  It is a fundamental observation of \cite{casiniarxiv} that the presence of generalized symmetries, such as our homotopy charge operator, is equivalent to the non-completeness of the algebras of local operators, which in turn, by a famous result of von Neumann's\cite{vn} and an implementation of Haag duality \cite{casiniarxiv}, is equivalent to the presence of non-locally generated operators. In general, in QFT, it is possible that such operators may become local when considered in a larger region. When both zeros and or poles are present, such operator localization is precluded. Consequently, in the IR, these operators have to be non-local, at least if they are relevant in the RG sense. Physically, if the Luttinger count does not equal the physical charge, then necessarily the propagating degrees of freedom are non-local.

\bibliography{reference} 

\section{Appendix: Particle Density}

By definition, the occupancy at a given momentum $k$  satisfies
\beq
\begin{split}
    \langle c_k^\dagger c_k\rangle&=-\langle Tc_k(0^-)c_k^\dagger(0) \rangle\\
    &=G(\tau=0^-,k)\\
    &=\frac{1}{\beta}\sum_ne^{-i\omega_n0^-}G(i\omega_n,k),
\end{split}
\eeq
where on the second line we have used the definition of the imaginary time Green function. Only at the end of our computation will we take the zero-temperature limit, $\beta\rightarrow\infty$. 
The summation over the Matsubara frequencies can be calculated using the contour integral method, 
\begin{equation}
\begin{split}
    \frac{1}{\beta}\sum_ne^{-i\omega_n0^-}G(i\omega_n,k)&=\oint_\Gamma\frac{dz}{2\pi i}n_F(z)e^{z0^+}G(z,k),
\end{split}
\label{eq:imcontour}
\end{equation}
where the contour $\Gamma$ goes around the imaginary axis clockwise, and $n_F(z)=1/(1+e^{\beta z})$ is the Fermi distribution function. For the Green function given by Eq. \eqref{eq:green2p}, since the integrand lacks poles except on the imaginary or real axis, we can deform the contour $\Gamma$ into a path counter-clockwise enclosing the real axis,
\beq
    \langle c_k^\dagger c_k\rangle&=&\oint_\Gamma\frac{dz}{2\pi i}n_F(z)e^{z0^+}G(z,k)\\
    &=&\int_{-\infty-i\epsilon}^{+\infty-i\epsilon}\frac{dz}{2\pi i}n_F(z)e^{z0^+}G(z,k)\nonumber\\
    &-&\int_{-\infty+i\epsilon}^{+\infty+i\epsilon}\frac{dz}{2\pi i}n_F(z)e^{z0^+}G(z,k)\\
    &=&Z_1n_F(\xi+U_1)+Z_2n_F(\xi+U_2).
\eeq
On the other hand, by taking the derivative on both sides of the identity $\Sigma(z,k)=G_0^{-1}(z,k)-G^{-1}(z,k)$, 
\beq
    \frac{\partial G^{-1}(z,k)}{\partial z}+\frac{\partial\Sigma(z,k)}{\partial z}=1
\eeq
reduces the density to
\beq
\oint_\Gamma\frac{dz}{2\pi i}n_F(z)e^{z0^+}G(z,k)= I_1+I_2.
\eeq
We first consider $I_1$ by deforming the contour $\Gamma$ into a path counter-clockwise enclosing the real axis
\beq
    I_1&=&\int_{-\infty-i\epsilon}^{+\infty-i\epsilon}\frac{dz}{2\pi i}n_F(z)e^{z0^+}\frac{\partial \ln G^{-1}(z,k)}{\partial z}\nonumber\\
    &-&\int_{-\infty+i\epsilon}^{+\infty+i\epsilon}\frac{dz}{2\pi i}n_F(z)e^{z0^+}\frac{\partial \ln G^{-1}(z,k)}{\partial z}.
\eeq
In the zero temperature limit, the Fermi distribution function vanishes on the right half plane and is equal to unity on the left half plane. Thus,
\begin{equation}
\begin{split}
    I_1&=\int_{-\infty-i\epsilon}^{-i\epsilon}\frac{dz}{2\pi i}\frac{\partial \ln G^{-1}(z,k)}{\partial z} -\int_{-\infty+i\epsilon}^{+i\epsilon}\frac{dz}{2\pi i}\frac{\partial \ln G^{-1}(z,k)}{\partial z}\\
    &=\frac{1}{2\pi i}\left.\ln\frac{G^{-1}(z-i\epsilon,k)}{G^{-1}(z+i\epsilon,k)}\right|_{-\infty}^0\\
    &=\left\{\begin{array}{cc}
        1 &  \mathrm{Re} G(0,k)>0\\
        0 & \mathrm{Re} G(0,k)<0 
    \end{array} \right.
\end{split}
\end{equation}
The Green function has an asymptotic form $G(z\rightarrow\infty)\approx1/z$. Thus, the real part of Green function is positive {\it iff} there are an odd number of poles or zeros located on the left half plane so that the sign of $G(z)$ changes from negative to positive. 
In order to compute $I_2$, we can further separate the integral into two terms using the explicit form of $G(z,k)$
\begin{equation}
\begin{split}
    I_2&=\oint_\Gamma\frac{dz}{2\pi i}n_F(z)e^{z0^+}G(z,k)\frac{\partial \Sigma(z,k)}{\partial z}\\
    &=\oint_\Gamma\frac{dz}{2\pi i}n_F(z)e^{z0^+}\left(\sum_i\frac{Z_i}{z-\xi(k)-U_i(k)}\right)\frac{\partial \Sigma(z,k)}{\partial z}\\
    &=I_{21}+I_{22}.
\end{split}
\end{equation}
Using the residue theorem, we get
\begin{equation}   
\begin{split}
I_{21}&=-Z_2n_F(\xi+U_1)+Z_2a_{12},\\ 
I_{22}&=-Z_1n_F(\xi+U_2)+Z_1a_{12},\\  
\end{split} 
\end{equation}
\end{document}